\documentclass[11pt,preprint,authoryear,longnamesfirst]{aastex}
\usepackage{psfig}
\received{2001 March~16}
\revised{2001}
\accepted{2001 June~21}
\journalid{560}{2001 October~20}
\paperid{53664}
\cpright{PD}{2001}
\shorttitle{Angular Wander of B2~2050$+$36}
\shortauthors{Lazio \& Fey}

\newcommand{\thd}{\ensuremath{\theta_d}}
\newcommand{\thr}{\ensuremath{\theta_r}}
\newcommand{\ld}{\ensuremath{l_d}}
\newcommand{\dtr}{\ensuremath{\Delta t_r}}
\newcommand{\thc}{\ensuremath{\theta_{\mathrm{cross}}}}
\newcommand{\thsep}{\ensuremath{\theta_{\mathrm{sep}}}}
\newcommand{\src}{\mbox{\protect\objectname[B2]{B2~2050$+$36}}}
\newcommand{\aips}{\textsc{aips}}
\newcommand{\cnsq}{\ensuremath{C_n^2}}
\newcommand{\kms}{\mbox{km~s${}^{-1}$}}
\newcommand{\mjybm}{\mbox{mJy~beam${}^{-1}$}}
\newcommand{\smunit}{\mbox{kpc~m${}^{-20/3}$}}

\begin{document}
\title{Multi-frequency VLBA Observations of the Compact Double B2~2050$+$36:  Constraints on Interstellar Scattering Revisited}

\author{T.~Joseph~W.~Lazio}
\affil{Code~7213, Remote Sensing Division, Naval Research Laboratory, 
	 Washington, DC 20375-5351  USA}
\email{lazio@rsd.nrl.navy.mil}

\and 
\author{A.~L.~Fey} 
\affil{US Naval Observatory, 3450 
        Massachusetts Ave.~NW, Washington, DC  20392-5420, USA}
\email{afey@usno.navy.mil}

\begin{abstract}
We present multi-frequency observations with the Very Long Baseline
Array of the compact double radio source \src.  Our observations are
at~0.33, 0.61, 1.67, 2.3, and~8.4~GHz, with the 0.61~GHz observations
forming the third epoch of observation of this source at that
frequency.  At~0.61~GHz, the structure of \src\ is dominated by two
components 56~mas apart.  Within the uncertainties of the various
measurements, this separation has remained unchanged for the past
16~years.  Any differential image wander caused by refractive
interstellar scattering is less than 4~mas.  Both the lack of
differential image wander and the frequency dependence of the angular
diameter of \src\ below~1~GHz indicate that the electron density power
spectrum along this line of sight has a spectral index near the
Kolmogorov value, with a value of~4 being highly unlikely.  We
conclude that diffractive scattering dominates along this line of
sight; results in the literature indicate that this conclusion also
holds true for the line of sight to the pulsar
\objectname[PSR]{PSR~B2020$+$28} (8\fdg7 from \src).  Comparison of
our 1.67~GHz observations with those obtained 21~years previously
place a limit on the projected linear separation velocity of the two
components of~$c$.
\end{abstract}


\section{Introduction}\label{sec:intro}

Density fluctuations on AU-size scales in the interstellar plasma may
produce a host of observable effects including refractive intensity
scintillations and image wander.  The magnitude of these effects
depends crucially upon the spectrum of the interstellar density
fluctuations.  If the spectrum is a power law with a spectral index
less than 4, these refractive effects will generally be masked by
diffractive effects from even smaller scale density fluctuations
\citep{rnb86}.  Conversely, if the spectrum is a power law with a
spectral index greater than 4, refractive effects will dominate.  For
many lines of sight the density spectral index appears to be close to,
but less than, 4 \citep{ars95}, with important exceptions
\citep[e.g.,][]{hwg85,cfc93,grl94,sfm96,rlg97}.

Complicating the effort to measure refractive effects is their
generally long time scale.  The relevant time scale is given by the
time it takes for an AU-sized density fluctuation to drift past the
observer.  For observations of pulsars, whose velocities can exceed
100~\kms, the time scale is days.  For observations of extragalactic
sources, the relevant velocity is a combination of the Earth's
velocity and random interstellar motions and is more likely to be
roughly 25~\kms, meaning that the refractive time scale is months to
years.

The source \src\ is an ideal source for probing refractive effects.
It lies on the outskirts of the Cygnus superbubble \citep{bs85}, a
region known to exhibit enhanced interstellar scattering
\citep{fsm89,lsc90,fsc91,wns94,mmrj95,sc98}.  VLBI observations reveal
its milliarcsecond structure to be dominated by two components
separated by approximately 60~mas, thus a search for refractive image
wander requires only \emph{relative} position measurements.

\citet[hereafter \citeauthor{fm93}]{fm93} obtained 0.61~GHz VLBI
observations of \src.  Combined with previous measurements in the
literature, they showed that scattering is moderately strong along
this line of sight.  They were also able to constrain the amount of
differential refractive image wander between the two components.
Their constraint, however, is based on measurements of the separation
of the components at two epochs separated by four years.  If the
refractive time scale is longer than four years, their constraint is
weakened considerably.  Assuming that typical interstellar velocities
along this line of sight are 50--100~\kms, \citeauthor{fm93} argued that the
refractive time scale is comparable to or less than four years.
Obviously, if the typical interstellar velocity is more like 25~\kms,
the refractive time scale is much longer.

We have acquired a third epoch of 0.61~GHz VLBI observations of \src,
specifically to confront the conclusions of \citeauthor{fm93} with
another epoch of observations.  This third epoch is 12 to~16~yrs after
the first two epochs and should allow us to place much better
constraints on any refractive wander.  We have also acquired
simultaneous or quasi-simultaneous observations at a range of other
wavelengths in order to confirm previous measurements of the strength
of interstellar scattering along this line of sight.

In \S\ref{sec:observe} we describe our observations and data analysis,
in \S\ref{sec:scatter} we discuss the implications of our observations
for both diffractive and refractive interstellar scattering along this
line of sight, and in \S\ref{sec:intrinsic} we discuss limits our
observations place on the intrinsic structure and kinematics of the
source.  We present our conclusions in \S\ref{sec:conclude}.

\section{Observations and Data Analysis}\label{sec:observe}

We observed \src\ with the VLBA for a total of~10~hours on~2000
July~25.  Relevant observation parameters are summarized in
Table~\ref{tab:log}.

\begin{deluxetable}{cccccc}
\tablewidth{0pc}
\tabletypesize{\small}
\tablecaption{VLBA Observing Log\label{tab:log}}
\tablehead{
	& & \colhead{Recorded} & \colhead{On-source} &
	\colhead{Synthesized} & \colhead{Image} \\
	\colhead{Frequency} & \colhead{Bandwidth} & \colhead{Polarization} 
	& \colhead{Time} & \colhead{Beam} & \colhead{Noise Level\tablenotemark{a}} \\
	\colhead{(GHz)} & \colhead{(MHz)} & & \colhead{(hr)} &
	\colhead{(mas)} & \colhead{(\mjybm)}
}

\startdata

0.33 & 12    & R,L & 2.88\tablenotemark{b} & 61 $\times$ 58 @ 87\arcdeg       & 3.2 \\
0.61 & \phn4 & R,L & 2.88\tablenotemark{b} & 27 $\times$ 19 @ 14\arcdeg       & 5 \\
1.67 & 32    & R,L & 2.93                  & 7.3 $\times$ 4.7 @ $-7.9\arcdeg$ & 0.78 \\
2.27 & 16    & R,L & 2.92\tablenotemark{c} & 5.8 $\times$ 3.6 @ $-9.4\arcdeg$ & 1.1 \\
8.42 & 16    & R,L & 2.92\tablenotemark{c} & 1.9 $\times$ 1.4 @ $-10\arcdeg$  & 0.61 \\

\enddata

\tablenotetext{a}{All images are of Stokes~I polarization.}
\tablenotetext{b}{Observations at~0.33 and~0.61~GHz were conducted simultaneously.}
\tablenotetext{c}{Observations at~2.27 and~8.42~GHz were conducted simultaneously.}

\end{deluxetable}

The observation frequencies were 0.33, 0.61, 1.67, 2.3, and~8.4~GHz;
both right and left circular polarization were recorded at all
frequencies.  The 2.3 and~8.4~GHz observations were obtained
simultaneously in ``S/X mode.''  Simultaneous 0.33 and~0.61~GHz
observations also were obtained by setting two of the eight baseband
converters (BBCs) at each VLBA antenna to record the two senses of
circular polarization at~0.61~GHz.  Because of severe radio frequency
interference around 0.61~GHz, filters restricted the observable
bandwidth for each BBC to~4~MHz.  The remaining BBCs recorded
0.33~GHz.  Formatter problems at the Kitt Peak antenna meant that only
nine of the ten VLBA antennas were correlated.

The correlated visibilities were processed in the standard fashion
within the NRAO Astronomical Image Processing System (\aips).
Fringe-fitting intervals ranged from 10~min.\ at~1.67~GHz to~3~min.\
at~0.33 and~8.4~GHz.  At~0.33 and~0.61~GHz, substantial editing of the
calibrated visibilities was required to remove radio frequency
interference.

In general the calibrated visibilities were then exported to the
Caltech \texttt{difmap} program for hybrid mapping.  To ensure a
robust estimate for the separation of the components
(\S\ref{sec:ref}), the 0.61~GHz observations were imaged using both
\aips\ and \texttt{difmap}; the 0.33~GHz observations were imaged
using only \aips.

Figure~\ref{fig:images} shows the images of \src\ at the various
frequencies.  The observed rms noise level in these images ranges
from~0.61~\mjybm\ at~8.42~GHz to~5~\mjybm\ at~0.61~GHz and are
summarized in Table~\ref{tab:log}.  The rms noise levels in our images
are roughly within a factor of~2--3 of the expected value.

\begin{figure}
\begin{center}
\mbox{\psfig{file=LF_f1a.ps,width=0.46\textwidth,silent=} \psfig{file=LF_f1b.ps,width=0.46\textwidth,silent=}}
\vspace{-0.25cm}
\mbox{\psfig{file=LF_f1c.ps,width=0.46\textwidth,silent=} \psfig{file=LF_f1d.ps,width=0.46\textwidth,silent=}}
\end{center}
\vspace{-1.25cm}
\caption[B2_2050+36.??.IM.PS]{VLBI images of \src.
(\textit{a})~0.33~GHz.  The rms noise level is 3.2~\mjybm, and the
contour levels are 3.2~\mjybm\ $\times$~$-2$, 3, 5, 7.1, 10, \ldots.
The beam is 60.8~mas $\times$ 58.5~mas at a position angle
of~87\arcdeg\ and is shown in the lower left.
(\textit{b})~0.61~GHz.  The rms noise level is 5~\mjybm, and the
contour levels are 5~\mjybm\ $\times$~$-3$, 3, 5, 7.1, 10, \ldots.
The beam is 27~mas $\times$ 19~mas at a position angle
of~14\arcdeg\ and is shown in the lower left.
(\textit{c})~1.67~GHz.  The rms noise level is 0.78~\mjybm, and the
contour levels are 0.78~\mjybm\ $\times$~$-3$, 3, 5, 7.1, 10, \ldots.
The beam is 7.3~mas $\times$ 4.7~mas at a position angle
of~$-7.9\arcdeg$ and is shown in the lower left.
(\textit{d})~2.3~GHz.  The rms noise level is 1.1~\mjybm, and the
contour levels are 1.1~\mjybm\ $\times$~$-3$, 3, 5, 7.1, 10, \ldots.
The beam is 5.8~mas $\times$ 3.6~mas at a position angle
of~$-9.4\arcdeg$ and is shown in the lower left.}
\label{fig:images}
\end{figure}

\setcounter{figure}{0}
\begin{figure}
\begin{center}
\mbox{\psfig{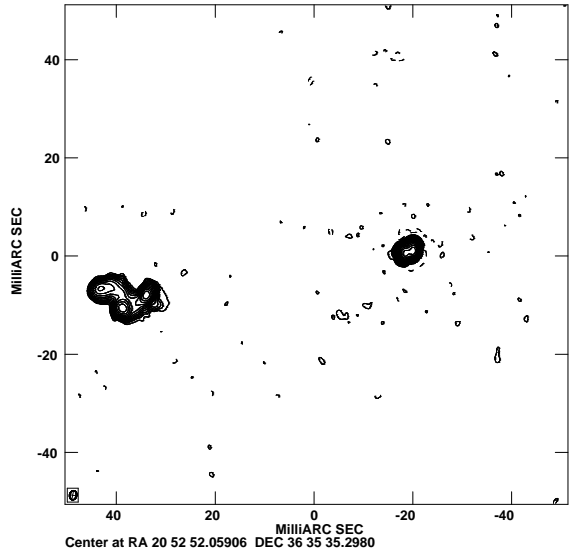}}
\end{center}
\vspace{-1.25cm}
\caption[B2_2050+36.??.IM.PS]{\textit{(Cont.)} VLBI images of \src.
(\textit{e})~8.4~GHz.  The rms noise level is 0.61~\mjybm, and the
contour levels are 0.61~\mjybm\ $\times$~$-3$, 3, 5, 7.1, 10, \ldots.
The beam is 1.9~mas $\times$ 1.4~mas at a position angle
of~$-10\arcdeg$ and is shown in the lower left.}
\end{figure}

Using the images shown in Figure~\ref{fig:images} as a guide, we fit
the self-calibrated visibility data with multiple gaussian component
models.  The number of gaussian components fitted ranged from~1
at~0.33~GHz to~7 at~8.42~GHz.  These models are summarized in
Table~\ref{tab:models}.  Below we shall make use of the fitted angular
diameters.  At the higher frequencies, we take the uncertainty in a
fitted angular diameter to be one half of the synthesized beam; at the
lower frequencies, where a one- or two-component fit suffices, we take
the uncertainties to be the square roots of the diagonal elements of the
covariance matrix obtained from a least-squares fit to the visibility
data (provided by the \texttt{UVFIT} fitting routine within \aips);
these values are of order 0.5~mas.

\begin{deluxetable}{lcccccccc}
\tablewidth{0pc}
\tablecaption{Gaussian Source Models\label{tab:models}}
\tablehead{
 \colhead{$\nu$} & \colhead{$S$} & \colhead{$r$} & \colhead{$\theta$} 
 & \colhead{$a$} & \colhead{$b/a$} & \colhead{$\phi$} 
 & \colhead{Component} \\
 \colhead{(GHz)} & \colhead{(Jy)} & \colhead{(mas)} & \colhead{(\arcdeg)}
 & \colhead{(mas)} &               & \colhead{(\arcdeg)}
}

\startdata
0.33 & 2.17 & 0.0 & \nodata & 106.2 & 0.76 & $-35$ & E\tablenotemark{a}\\

0.61 & 2.48 &  0.0 & \nodata & 24.1 & 0.82 & 24 & E\\
     & 0.39 & 56.1 &   $-82$ & 26.2 & 0.69 &  8 & W\\ 
     & 0.11 & 32.6 &       8 & 47.5 & 0.00 & 67 & E\\

1.67 & 1.75 & 59.3 &  $-80$ &  3.5 & 0.90 & $-49$ & W\\
     & 0.79 &  5.5 &  $-82$ &  4.2 & 0.74 & $-19$ & E \\
     & 0.99 &  1.3 & $-132$ &  4.0 & 0.86 &   33  & E \\
     & 1.06 &  3.4 &    42  &  4.9 & 0.69 & $-77$ & E\\
     & 0.16 &  8.9 & $-106$ & 11.8 & 0.32 & $-52$ & E\\
     
2.27 & 1.97 &  0.0 & $-32$ & 2.9 & 0.71 & $-61$ & W\\
     & 0.92 & 58.5 &   101 & 3.7 & 0.81 &    17 & E\\
     & 0.58 & 53.8 &   100 & 3.3 & 0.59 & $-25$ & E\\
     & 0.83 & 61.5 &    97 & 4.2 & 0.54 & $-78$ & E\\
     & 0.12 & 52.9 &   103 & 7.5 & 0.39 & $-46$ & E\\

8.42 & 0.35 & 18.4 & $-89$ & 0.8 & 0.54 &   84 & W\\
     & 0.39 & 20.1 & $-87$ & 1.0 & 0.51 &  $-1$& W \\
     & 0.19 & 34.9 &  103  & 2.1 & 0.64 & $-29$& E \\
     & 0.19 & 40.3 &  105  & 2.1 & 0.54 &   60 & E \\
     & 0.19 & 43.2 &   99  & 3.5 & 0.51 & $-85$& E \\
     & 0.22 & 18.8 & $-85$ & 3.1 & 0.24 & $-67$& W \\
     & 0.16 & 39.0 &  102.4& 6.3 & 0.28 &  40  & E \\
\enddata

\tablecomments{The models fitted to the visibility data are Gaussians
with flux density~$S$ and FWHM major axis~$a$ and minor axis~$b$, with
major axis in position angle~$\phi$ (measured north through
east). Components are separated from the (arbitrary) origin of the
image by a distance~$r$ in position angle~$\theta$, which is the
position angle (measured north through east) of a line joining the
components with the origin.  Components and sub-components are
identified as with either an ``E'' or ``W'' corresponding to whether
they are Eastern or Western, respectively.}
\tablenotetext{a}{Assumed Eastern component; see \S\ref{sec:diff}.}
\end{deluxetable}

\section{Constraints on Interstellar Scattering}\label{sec:scatter}

We now revisit the scattering properties of the line of sight to \src.
We consider both diffractive effects, namely angular broadening, and
refractive effects, namely differential image wander and refractive
intensity fluctuations.

We shall make the standard assumption throughout that the density
fluctuations responsible for interstellar scattering can be
parameterized by an exponentially-truncated power law
\begin{equation}
P_{\delta n_e}(z, q) = \cnsq(z)\,q^{-\alpha}e^{-(q/q_1)^2}, 
 \quad q > q_0.
\label{eqn:spectrum}
\end{equation}
Here $q_1$ is the largest wavenumber on which density fluctuations
occur, which is related to the smallest length scales for density
fluctuations~$l_1$ or ``inner scale'' as $q_1 = 2\pi/l_1$.  The
smallest wavenumber on which density fluctuations occur or the ``outer
scale'' is $q_0 = 2\pi/l_0$.  The constant~\cnsq\ is a measure of the
level of density fluctuations and is related to the rms density as
$\langle n_e^2\rangle = [2(2\pi)^{4-\alpha}/(\alpha-3)]\,\cnsq\,
l_0^{3-\alpha}$ \citep{crwfs91}.  The power law spectral index is
assumed to provide clues about the process(es) that generate or
maintain the density fluctuations.  Our observations allow us to
measure $\alpha$, constrain \cnsq, and indirectly place limits on
$l_1$.

\subsection{Diffractive Scattering}\label{sec:diff}

\subsubsection{Angular Broadening and Density Spectrum}\label{sec:broaden}

An infinitely-distant point source viewed through a medium filled with
density fluctuations having a spectrum described by
equation~(\ref{eqn:spectrum}) and a spectral index $\alpha = 11/3$
will be broadened to an angular diameter \citep{cl91,tc93}
\begin{equation}
\thd = \cases{
  128\,\mathrm{mas}\>\mathrm{SM}^{3/5}\nu_{\mathrm{GHz}}^{-11/5} & $\thd \lesssim \thc$; \cr
  132\,\mathrm{mas}\>\mathrm{SM}^{1/2}\nu_{\mathrm{GHz}}^{-2}\left(\frac{l_1}{100\,\mathrm{km}}\right)^{-1/6} & $\thd \gtrsim \thc$. \cr}
\label{eqn:thd}
\end{equation}
Here $\nu_{\mathrm{GHz}}$ is the observing frequency measured in GHz,
and the scattering measure~SM is the line of sight integral of \cnsq, 
\begin{equation}
\mathrm{SM} = \int dz\,\cnsq(z).
\label{eqn:sm}
\end{equation}
The quantity \thc\ is the cross-over angle and describes the relative
importance of the inner scale~$l_1$.  It is given by $\thc \simeq
0\farcs16\,\nu_{\mathrm{GHz}}^{-1}(l_1/100\,\mathrm{km})^{-1}$.  The
constants (128~mas and 132~mas) given in equation~(\ref{eqn:thd}) are
those appropriate for infinitely distant sources and differ slightly
from those given by \cite{cl91} who considered sources embedded in the
medium.  These formulae specify the diameter as the full width at half
maximum (FWHM) of the shape of the image and that shall be the
convention we adopt in reporting the measured diameters.

The scattering diameter~\thd\ defines the diffractive scale~\ld, $\ld
= \lambda/\thd$ (at observing wavelength~$\lambda$).  If the typical
interferometer baseline~$b$ is sufficient to resolve \thd\ (i.e., $b
\sim \ld$) and $b \lesssim l_1$, we expect $\thd \gtrsim \thc$ and
$\thd \propto \nu^{-2}$.  Conversely if $b \gtrsim l_1$, then we
expect $\thd \lesssim \thc$ and $\thd \propto \nu^{-11/5}$.  Therefore
fitting the observed diameter as a function of frequency can constrain
both $l_1$ and $\alpha$.

There are alternate causes for a $\thd \propto \nu^{-2}$ dependence.
For instance, if the spectrum of density fluctuations is not a power
law but a gaussian centered on a particular length scale, that can
produce a $\thd \propto \nu^{-2}$ dependence.  Similarly, if the
density spectrum is dominated just slightly by AU-size scales such
that $\alpha \gtrsim 4$, that could also produce an apparent $\thd
\propto \nu^{-2}$ dependence within the measurement uncertainties.  As
we shall show below, $\alpha < 4$ so that we could interpret a $\thd
\propto \nu^{-2}$ dependence as resulting from a ``large'' inner
scale.

\begin{figure}
\begin{center}
\mbox{\psfig{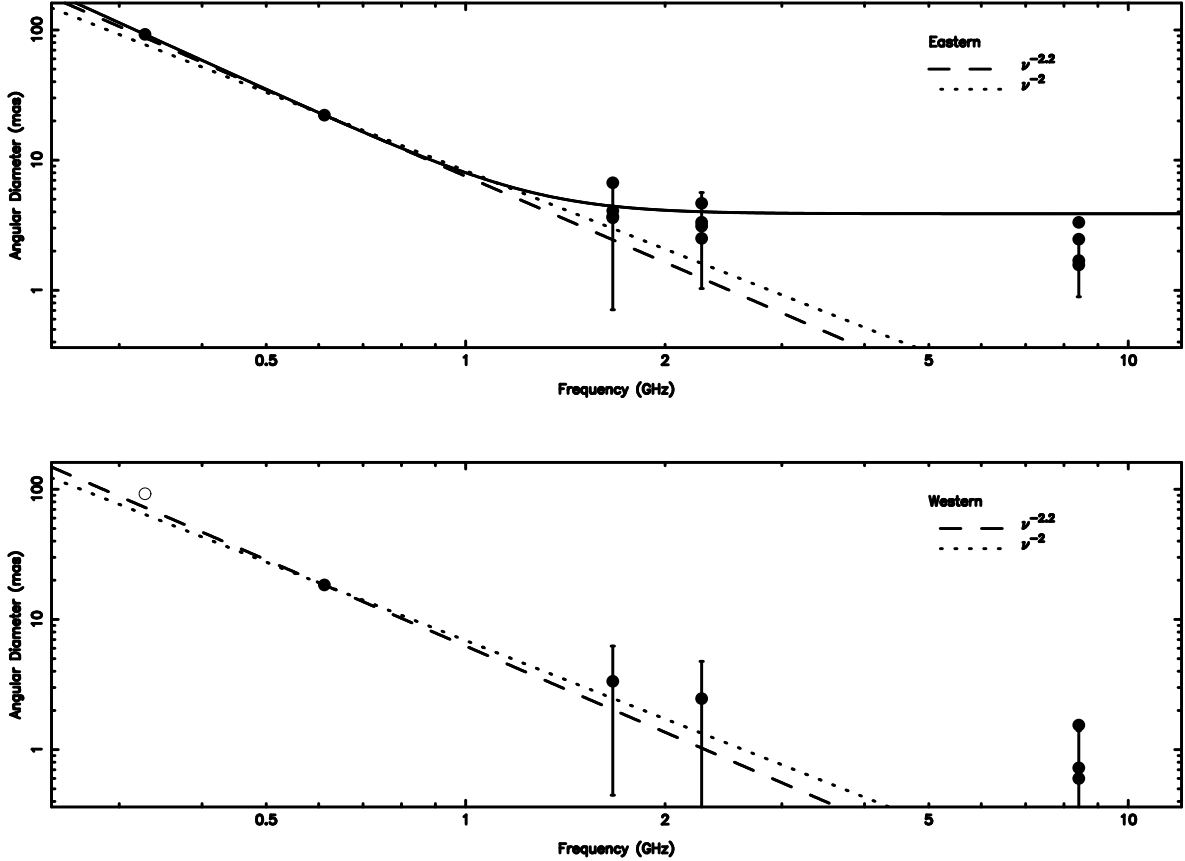}}
\end{center}
\vspace{-0.5cm}
\caption[diameter.ps]{The angular diameter of the components of
\src\ as a function of observing frequency.  All components listed in
Table~\ref{tab:models} are plotted.
\textit{Top}: The eastern component.
\textit{Bottom}: The western component.  For the western component we
also show, as an open circle, the diameter it would have assuming that
its scattering diameter is identical to that of the eastern
component.  This point is shown for display purposes only and is not
used in any of the fits described.  
In both panels a dashed line shows the dependence $\theta \propto
\nu^{-2.2}$, as expected for a Kolmogorov spectrum, and a dotted line
shows $\theta \propto \nu^{-2}$.  Both lines are normalized to pass
through the measured angular diameter at~0.61~GHz.  For the eastern
component only, a solid line shows the fit, which allows for intrinsic
structure and which is discussed in the text.  At the higher
frequencies, we plot all sub-components, but we show the uncertainty on
only one of them as we assume the uncertainty to be the same for all
sub-components and equal to one half of the synthesized
beam.}
\label{fig:diameter}
\end{figure}

Figure~\ref{fig:diameter} shows the measured angular diameters as a
function of observing frequency.  Also shown are two curves, one for
$\theta \propto \nu^{-2.2}$ appropriate for a spectrum with $\alpha =
11/3$ and a diffractive scale~\ld\ much larger than $l_1$ and one for
$\theta \propto \nu^{-2}$ appropriate for the case in which $\ld
\lesssim l_1$.  

As Figure~\ref{fig:images} demonstrates, \src\ is not only not a point
source, but the two components of it can be resolved into multiple
sub-components at high frequencies.  A visual inspection of
Figure~\ref{fig:diameter} suggests that at frequencies below~1~GHz
scattering dominates the structure of \src\ while at frequencies
above~1~GHz there are clear deviations from either curve due
presumably to intrinsic structure.  In order to quantify this result,
we have fit the measured diameters of the eastern component
(Table~\ref{tab:models}) with the following functional form
\begin{equation}
\theta^2
 = \theta_i^2 + \theta_s^2\nu_{\mathrm{GHz}}^{-2\alpha/(\alpha-2)}.
\label{eqn:fit}
\end{equation}
Here $\theta_i$ is the intrinsic diameter and $\theta_s$ is the
scattering diameter, both at the fiducial frequency of~1~GHz.  Unlike
in equation~(\ref{eqn:thd}) where we specified the dependence for the
specific case of $\alpha = 11/3$, here we allow the frequency
dependence to be part of the fit.  The frequency dependence we use is
appropriate for $\alpha < 4$.

Several points are in order regarding the choice of data we used.
At~0.33~GHz only one component is seen
(Figure~\ref{fig:images}\textit{a}).  Attempts to fit the visibility
data with a two-component structure suggested by the 0.61~GHz image
produced a poorer fit to the data.  We have subtracted a one-component
model (Table~\ref{tab:models}) from the visibility data and imaged the
residuals.  There is no indication of a second component at the
location expected from the higher frequency images.  We have also
produced model visibility data, comprised of two components whose flux
densities, diameters, and positions are consistent with those seen at
higher frequencies.  Fitting these model data with a one-component
model results in a fit nearly the same as what we find by fitting the
observed visibility data.  The similarity suggests that any systematic
error in determining the diameter is probably no more than 1~mas.
Because of the self-calibration we performed, we have no absolute
position information.  Given that the eastern component is stronger
at~0.61~GHz, we have therefore assumed that the single component seen
at~0.33~GHz is the eastern component.

At the higher frequencies both components can be resolved into
multiple sub-components.  In Figure~\ref{fig:diameter} we plot the
diameters of the individual sub-components.  The diameters of the
individual sub-components are clearly not representative of the
diameter of the entire component.  Therefore, as a measure of the size
of the components at these frequencies, we have added the diameters of
the sub-components (Table~\ref{tab:models}) in quadrature.  If
anything, this estimate will \emph{underestimate} the size of a
component.  However, our primary interest is not in the intrinsic
structure, and this procedure suffices to obtain an estimate for the
intrinsic size.

\src\ is classified as a gigahertz-peaked spectrum source, and both
components show a single flux density maximum near~1~GHz
\citep{mhp85}.  Accordingly, in equation~(\ref{eqn:fit}) we have
assumed that the intrinsic structure has no frequency dependence.
While we have not conducted extensive tests, assuming no frequency
dependence produces clearly much better fits than does assuming a
frequency dependence of $\nu^{-1}$, as might be expected for a
homogeneous source with a peak brightness temperature~$T_B$
\citep{ko88}.

We have used a grid-search technique to minimize the difference, in a
$\chi^2$ sense, between the data and the expression of
equation~(\ref{eqn:fit}) for the three parameters $\theta_i$,
$\theta_s$, and~$\alpha$.  We have also determined the allowed ranges
for the parameters in the following manner:  Holding two parameters
fixed at the values that minimize $\chi^2$, we change the third
parameter until $\chi^2$ changes by unity.

Table~\ref{tab:fit} summarizes the best fit value for the parameters
and their allowed ranges.  The best-fitting curve is also shown in
Figure~\ref{fig:diameter}.  A key result of this fit is that a density
spectrum spectral index of~4 is highly unlikely.  This result means
that the inner scale~$l_1$ is substantially smaller than most
baselines in the array or that $\thd < \thc$.  The typical array
baseline is roughly 2000~km, so we place an upper limit of $l_1 \ll
2000$~km.  This constraint is not a serious challenge to the value
estimated by \cite{sg88} of $l_1 \lesssim 200$~km.  Using
equation~(\ref{eqn:thd}) we can also derive the scattering
measure~\hbox{SM}. Table~\ref{tab:fit} also lists the values of these
derived quantities.

\begin{deluxetable}{ccc}
\tablecolumns{3}
\tablewidth{0pc}
\tablecaption{Best-Fit Diffractive Scattering Model Parameters for
Eastern Component at~1~GHz\label{tab:fit}}
\tablehead{
 \colhead{ }         & \colhead{Adopted} & \colhead{} \\
 \colhead{Parameter} & \colhead{Value}   & \colhead{Range}
}
\startdata
$\theta_i$ & 3.9~mas & 3.5--4.3~mas \\
$\theta_s$ & 7.0~mas & 6.98--7.03~mas \\
$\alpha$   & 3.52    & 3.51--3.53 \\

\cutinhead{Derived Scattering Parameters}
SM\tablenotemark{a} & $10^{-2.1}$~\smunit \\
$l_1$               & $\ll 2000$~km \\
\enddata
\tablenotetext{a}{Kolmogorov density spectrum, $\alpha = 11/3$,
assumed.}
\end{deluxetable}

\subsubsection{Image Anisotropy and Orientation}\label{sec:shape}

The anisotropy and orientation of a scattered image can, in principle,
yield information about the density structures responsible for the
angular broadening \citep[e.g.,][]{wns94,tmr98,cl01}.  The
simultaneous low frequency observations we report here are seemingly
ideal for constraining any wavelength-dependent changes in the
anisotropy or orientation of \src, as the $u$-$v$ plane coverage
should be identical (once scaled by the ratio of the wavelengths).

We consider first the anisotropy of the image.  We find axial ratios
of $b/a \simeq 0.80$ (Table~\ref{tab:models}) or, in terms of the
image anisotropy parameter of \cite{rnb86}, $e_s \simeq
0.25$. \cite{rnb86} also provide an estimate of the image anisotropy to
be expected from random fluctuations within the medium.  For our
observing wavelengths and scattering parameters derived above, the
expected random contribution to $e_s$ is approximately 0.03.  We
conclude that these image anisotropies are significant.  These
anisotropies are also comparable to those of other sources seen
through the Cygnus region \citep{sc88,wns94,mmrj95,sc98,df01}.
\cite{df01} attribute the similarity of the axial ratios, between
different sources at different frequencies and different epochs, to be
indicative of an anisotropy in the scattering material itself.  

We cannot conclude that the image anisotropy changes with frequency
for \src.  The formal uncertainties from the fitting of the visibility
data are approximately 1\%.  Even this small uncertainty on the
angular diameters, when propagated to the axial ratios, implies an
uncertainty of $\Delta(b/a) \simeq 0.03$, comparable to the difference
between the axial ratios at the two frequencies
(Table~\ref{tab:models}).

Changes in the image orientation between the two frequencies may
reflect changes in the orientation of the density irregularities
responsible for the scattering.  Unfortunately, we also cannot draw
any conclusions about the image orientation.  Although the $u$-$v$
plane coverage should be similar, differences in the editing (due to
the significant levels of RFI encountered) change the relative $u$-$v$
plane coverage significantly.  An indication of the difference can be
seen by inspecting the synthesized beams shown in
Figures~\ref{fig:images}\textit{a} and~\ref{fig:images}\textit{b}.  In
particular, the synthesized beam at~610~MHz has an orientation
of~14\arcdeg compared to an orientation for the eastern component
of~24\arcdeg.  Thus, the apparent change in orientation
between~330~MHz ($-35\arcdeg$) and~610~MHz ($24\arcdeg$) may be due
largely to the differences in $u$-$v$ coverage.

\subsection{Refractive Interstellar Scattering}\label{sec:ref}

Large scale density fluctuations may affect the image position and
flux density.  In this section we place constraints on the strength of
refractive modulation of the image, focussing primarily on the
0.61~MHz image.

\subsubsection{Differential Image Wander}\label{sec:wander}

Density fluctuations on a scale $l_r = D\thd$, where $D$ is the
characteristic distance to the scattering medium, will cause
refractive ``bending'' and ``steering'' of the propagating rays.  In
turn, this will affect the angle of arrival of rays and make radio
sources wander about a true or nominal position.  The magnitude of
this effect depends both upon the strength of scattering and the value
of $\alpha$.

\citeauthor{fm93} selected \src\ because its two component structure
allowed for \emph{relative} position measurements to be performed,
obviating the need for determining absolute positions with
milliarcsecond accuracy.  \citeauthor{fm93} summarized the predictions
for the amount of differential angular wander as a function of
$\alpha$ as predicted by various groups (\citealt{cpl86,rnb86,rc88};
see Appendix~A of
\citeauthor{fm93}).  They found that the rms differential refractive
angular wander is
\begin{equation}
\langle\Delta\thr^2\rangle^{1/2} = \cases{
  F(\alpha)\left(\frac{D}{\lambda}\right)^{(\alpha-4)/2}\thd^{\alpha-3} & $\alpha < 4$; \cr
  G(\alpha)\left(\frac{r_{\mathrm{sep}}}{l_r}\right)^{(\alpha-4)/2}\thd & $\alpha > 4$.
}
\label{eqn:diffrw}
\end{equation}
Here $r_{\mathrm{sep}}$ is the separation between pierce points
through the scattering medium.  For a double source (like \src) with a
component angular separation of $\thsep$, $r_{\mathrm{sep}} =
D\thsep$.  The functions $F(\alpha)$ and $G(\alpha)$ are normalization
functions that differ from group to group, but typically by no more
than a factor of a few.  For reference, \cite{cpl86} predict
$\langle\Delta\thr^2\rangle^{1/2} =
0.18\,\mathrm{mas}\,(D_{\mathrm{kpc}}/\lambda_{\mathrm{cm}})^{-1/6}\thd^{2/3}$
for a Kolmogorov spectrum in a screen located $D_{\mathrm{kpc}}$~kpc
distant, producing an angular broadening of \thd~mas, and observed at
a wavelength of $\lambda_{\mathrm{cm}}$~cm.

In general if $\alpha > 4$ refractive effects like angular wander
dominate over diffractive effects like angular broadening.  Thus one
would expect to see the apparent position of a source wander by more
than the width of its scattering disk.  If $\alpha < 4$, the converse
is true.

Table~\ref{tab:sep} summarizes the 0.61~MHz measurements of the
angular separation between the two components of \src.  We have
estimated $\langle\Delta\thr^2\rangle^{1/2}$ using two different
methods.  Henceforth we shall also distinguish between the ensemble
average expected differential angular wander,
$\langle\Delta\thr^2\rangle^{1/2}$, and the quantity estimated from
the separation measurements, $(\bar{\thr^2})^{1/2}$.  First, we
calculated $(\bar{\thr^2})^{1/2}$ using the standard estimator for the
variance and using the three values in Table~\ref{tab:sep}.  This
estimate is $(\bar{\thr^2})^{1/2} = 2.3$~mas.  Second, we estimated
$(\bar{\thr^2})^{1/2}$ to be comparable to the maximum difference
between component separations at the different epochs,
$\max(\Delta\theta)$.  This estimate is $\max(\Delta\theta) = 4$~mas.

\begin{deluxetable}{lcc}
\tablecolumns{3}
\tablewidth{0pc}
\tablecaption{\src\ Component Separation Measurements\label{tab:sep}}
\tablehead{
 \colhead{Epoch} & \colhead{\thsep} & \colhead{Reference} \\
                 & \colhead{(mas)}
}
\startdata
1984~August & 60 $\pm$ 3     & \citealt{mh86} \\    
1988~June   & 56 $\pm$ 2     & \citeauthor{fm93} \\ 
2000~July   & 56.1 $\pm$ 1.0 & this work \\         

\cutinhead{Differential Angular Wander Upper Limits}
$(\bar{\thr^2})^{1/2}$ & 2.3~mas \\
$\max(\Delta\theta)$  & 4~mas
\enddata
\end{deluxetable}

\begin{figure}
\begin{center}
\mbox{\psfig{file=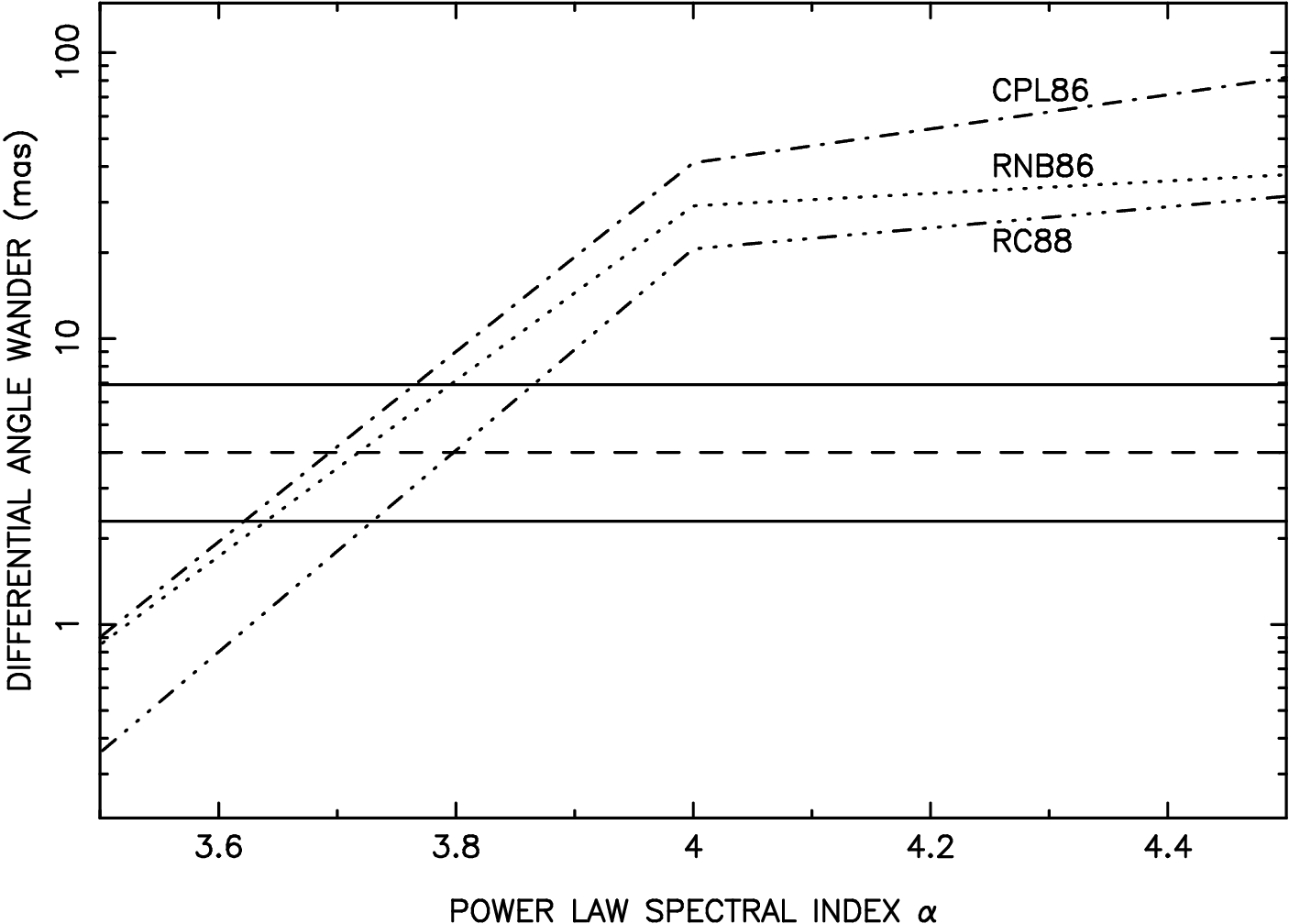,width=0.9\textwidth,angle=-90,silent=}}
\end{center}
\vspace{-0.5cm}
\caption[wander.ps]{Differential angular
wander~$\langle\Delta\thr^2\rangle^{1/2}$ as a function of the density
spectral index~$\alpha$.  Three expressions for
$\langle\Delta\thr^2\rangle^{1/2}$ are shown, that of \cite{cpl86}
(dot-dash line), \cite{rnb86} (dotted line), and \cite{rc88}
(dot-dot-dot-dash line).  All expressions are taken from Appendix~A of
\citeauthor{fm93}.  The abrupt change in slope at $\alpha = 4$ occurs
because no attempt has been made to join smoothly the expressions for
$\alpha < 4$ and $\alpha > 4$.  The solid horizontal lines represent
our upper limits on the differential angular wander of \src\ as given
by the standard variance estimator of $(\bar{\thr^2})^{1/2}$ and
$3(\bar{\thr^2})^{1/2}$.  The dashed horizontal line represents our
upper limit on the differential angular wander as given by
$\max(\Delta\theta)$.}
\label{fig:wander}
\end{figure}

Figure~\ref{fig:wander} shows $\langle\Delta\thr^2\rangle^{1/2}$ as a
function of the density spectral index~$\alpha$, as predicted by
\cite{cpl86}, \cite{rnb86}, and \cite{rc88}.  We also plot our
observational limits $(\bar{\thr^2})^{1/2}$, $3(\bar{\thr^2})^{1/2}$,
and $\max(\Delta\theta)$.  In producing Figure~\ref{fig:wander} we
have assumed $D = 2$~kpc.  We have considered
$\langle\Delta\thr^2\rangle^{1/2}$ as a function of both $D$
and~$\alpha$.  However, the dependence on $D$ is fairly weak,
$\langle\Delta\thr^2\rangle^{1/2} \propto D^{-1/6}$ for $\alpha =
11/3$.  Reasonable ranges for $D$ produce little
difference in $\langle\Delta\thr^2\rangle^{1/2}$.  The value $D =
2$~kpc is motivated by the distance to the Cygnus superbubble
\citep{bs85}, through which the line of sight to \src\ passes.

Figure~\ref{fig:wander} shows clearly that along this line of sight
the upper limits on $\langle\Delta\thr^2\rangle^{1/2}$ constrain
$\alpha < 4$.  The largest allowed value, $\alpha = 3.86$ occurs if we
adopt an upper limit of $3(\bar{\thr^2})^{1/2}$ and use the expression
for $\langle\Delta\thr^2\rangle^{1/2}$ from \cite{rc88}.  Using a more
stringent limit on the differential angular wander,
$\max(\Delta\theta)$, the expressions for
$\langle\Delta\thr^2\rangle^{1/2}$ from \cite{cpl86} or \cite{rnb86}
we find $\alpha \lesssim 3.7$.  We adopt $\alpha = 11/3$, as a limit
based on $(\bar{\thr^2})^{1/2}$ we consider to be overly strict.

The standard estimator for the variance, which we have used in
determining $(\bar{\thr^2})^{1/2}$, assumes that the measurements are
independent.  In this case the independence of the various
measurements can be assessed by comparing the interval between the
different measurement epochs with the refractive time scale, 
\begin{equation}
\dtr 
 = \frac{l_r}{v} 
 = 20\,\mathrm{yr}\left(\frac{D}{2\,\mathrm{kpc}}\right)\left(\frac{v}{10\,\kms}\right)^{-1},
\label{eqn:dtr}
\end{equation}
where $v$ is the bulk velocity of the material.  \citeauthor{fm93}
assumed $v = 50$--100~\kms, for which $\dtr = 2$--4~yr.  In this case,
all three measurement epochs (Table~\ref{tab:sep}) could be considered
independent, though the first two only marginally so.  However, if the
characteristic velocity of the scattering medium is smaller, by as
little as a factor of~2, the refractive time scale would be long
enough that the first two epochs (1984~August and 1988~June) could not
be considered independent.

If the characteristic velocity is as small as 10~\kms, then less than
one refractive time has elapsed since the first observation and none
of these observations can be considered to be independent.  Even if
this was the case, we would expect to see some change over the 12~yr
time span separating the second and third epochs (1988~June and
2000~July).  That essentially no change in the angular separation is
seen indicates that $\max(\Delta\theta)$ provides a reasonably
stringent upper limit on $\langle\Delta\thr^2\rangle^{1/2}$.

\subsubsection{Refractive Intensity Fluctuations}\label{sec:riss}

In concert with refractive wandering, the large-scale density
fluctuations will also produce intensity fluctuations as more or fewer
rays are ``steered'' into the line of sight.  In the case of \src,
this will produce independent variations in the flux density of its
two components.

\citeauthor{fm93} compared the flux densities of the two components
between their observations and those of \cite{mh86}.  They found that
the flux density of the components was consistent with no change in
the 4~yr between 1984~August and 1988~June.  Table~\ref{tab:flux}
summarizes these earlier measurements and those reported here.

\begin{deluxetable}{lcccc}
\tablewidth{0pc}
\tablecaption{Flux Density of \src\ at~0.61~GHz\label{tab:flux}}
\tablehead{
 \colhead{Epoch} & \colhead{Eastern} & \colhead{Western} &
 \colhead{Ratio} & \colhead{Reference} \\
                 & \colhead{(Jy)}    & \colhead{(Jy)}
}
\startdata
1984~August & 3.8               & 0.6               & 0.16            & \citealt{mh86} \\    
1988~June   & 3.9 $\pm$ 0.1     & 0.6 $\pm$ 0.1     & 0.15 $\pm$ 0.03 & \citeauthor{fm93} \\
2000~July   & 2.864 $\pm$ 0.006 & 0.374 $\pm$ 0.014 & 0.13 $\pm$ 0.01 & this work \\ 
\enddata
\end{deluxetable}

Because of changes in telescopes, receivers, correlators, and possible
intrinsic changes, we do not compare the absolute flux densities of
these earlier epochs to our measurements.  Rather we compare the flux
density ratio between the two components.  As Table~\ref{tab:flux}
shows, the western component has remained at a flux density of
approximately 15\% that of the eastern component for the past 15~yr.

\cite{gn85} present expressions for the rms refractive intensity
fluctuation~$m_r$ as a function of density spectral index~$\alpha$
\citep[see also][]{r86}.  In general, $m_r \sim 10$\% for
$\alpha < 4$ and $m_r \gtrsim 50$\% for $\alpha > 4$.  We determine an
observed value $m_r \sim 1.5$\% from the flux density ratios in
Table~\ref{tab:flux}, suggesting $\alpha < 4$.  For $\alpha < 4$, the
strong scattering, i.e., large SM, determined from the observed
scattering diameter (Table~\ref{tab:fit}) means that the rms
refractive intensity fluctuation will be only a slowly-changing
function of~$\alpha$.  From our observed value of~$m_r$ we find
\citep[][equation~3.1.5]{gn85} $\alpha < 3.85$, consistent with the
value found by fitting the angular diameters (Table~\ref{tab:fit}).

\subsubsection{Comparison with \protect\objectname[PSR]{PSR~B2020$+$28}}\label{sec:pulsar}

The pulsar \objectname[PSR]{PSR~B2020$+$28} (8\fdg7 from \src) has
been the subject of many monitoring programs designed at least in part
to measure refractive scattering along the line of sight to it.  In
all cases, it is found that diffractive scattering dominates
refractive scattering, as we find for the line of sight to \src.

\cite{gsw85} used dynamic spectra obtained at 0.41~MHz to measure the
frequency drift rate and estimated that the diffractive scattering
angle was at least 5 times that of the refractive scattering angle.
Using a structure function analysis of a 0.43~GHz flux density
monitoring program, \cite{lrc94} find diffractive scattering to
dominate over refractive scattering.  \cite{bgr99} monitored the
pulsar's flux density at~0.33~MHz and estimated the diffractive and
refractive scattering angles toward this pulsar.  They found that the
diffractive scattering angle was roughly 10 times larger than the
refractive scattering angle.  They also estimated a density spectrum
spectral index of $\alpha = 3.54$.  \cite{sc90} were unable to
observe any refractive modulations of the pulsar's flux density in
their multi-frequency monitoring program, consistent with diffractive
scattering dominating on this line of sight.

\section{Intrinsic Structure and Kinematics of
	\src}\label{sec:intrinsic}

A key assumption in our analysis of the differential image wander is
that there is little intrinsic change in the separation of the
components.  Given that on milliarcsecond scales, many sources show
evidence of jets and bulk motion, this assumption clearly must be justified.

We have compared our 1.67~GHz observations with those of \cite{pm81}.
This comparison is somewhat difficult because, in general, different
antennas at different locations were used in the two VLBI arrays.
Thus, this comparison is necessarily crude, but it will be sufficient
to show that our assumption is reasonable.

\cite{pm81} cite a separation of~60~mas between the two components
from observations in~1979 September.  If we fit our 1.67~GHz
observations with only two components, we find a separation of~58~mas,
\emph{less} than that of \cite{pm81}.  Even allowing for differences
in arrays and uncertainties in the fitting, we conclude that any
intrinsic change in the separation between the two components is
probably less than 1~mas.  Over the 20.8~yr separating these
measurements, any intrinsic motion is probably less than
0.05~mas~yr${}^{-1}$.  At a redshift of~0.354 for \src\ \citep{dobt00}
and for an assumed Hubble constant of~65~km~s${}^{-1}$~Mpc${}^{-1}$,
this proper motion corresponds to a projected linear separation
velocity of~$c$.  Our limit is similar to that found by \cite{skc99}
for \objectname[]{CTD~93}, a source with a similar compact double structure.

\section{Conclusions}\label{sec:conclude}

We have conducted a multi-frequency suite of observations on the
source \src.  We have used these observations to confirm and extend
previous conclusions, principally those of \citeauthor{fm93},
regarding interstellar scattering along this line of sight.

Our 0.61~MHz observations form the third epoch of VLBI observations at
this frequency.  Our observations are separated by~12 to~16 years from
the previous two epochs.  As do previous observations, we find the
structure of \src\ at this frequency to be dominated by two components
approximately 56~mas apart.  Within the uncertainties of the various
measurements, this separation has remained unchanged for the past
16~years.  We can place a reasonably strict limit of~4~mas on any
differential image wander caused by refractive interstellar
scattering.  In turn, this limit implies that the electron density
power spectrum along this line of sight has a spectral index near the
Kolmogorov value, with a value of~4 being highly unlikely.  As the
diffractive scattering diameter of the source at this frequency is
21~mas, we conclude that diffractive scattering dominates along this
line of sight.

The most significant source of uncertainty in this limit is probably
the velocity at which the scattering material is drifting past the
line of sight.  A ``high'' velocity of~100~\kms\ would mean that the
separations measured at the three epochs are independent.  A ``low''
velocity of~10~\kms\ would mean that none of the epochs are
independent.  Even if the velocity of the scattering material is low,
the separation between our observations and the previous ones is
probably long enough that, if refractive scattering were important on
this line of sight, some change in the apparent separation of the
components would be expected.  Nonetheless, future observations of
\src\ at~0.61~GHz or similar observations of other sources on lines of
sight for which the refractive time scale could be estimated would be
useful.

Our observations are in good agreement with a variety of observations
on the pulsar \objectname[PSR]{PSR~B2020$+$28} (line of sight 8\fdg7
from \src).  These observations suggest that diffractive scattering
also dominates refractive scattering along the line of sight to the
pulsar.

Our quasi-simultaneous observations range from~0.33 to 8.4~GHz.  We
have used these multi-frequency measurements to constrain the
frequency dependence of the angular diameter of one component of \src.
At frequencies below~1~GHz, the structure of the eastern component of
\src\ is dominated by scattering.  Fitting for both the density
spectrum spectral index~$\alpha$ and the scattering diameter~\thd, we
find $\alpha = 3.52$ with the value of~4 again highly unlikely.  At a
nominal frequency of~1~GHz, we find the scattering diameter to be
$\thd = 7$~mas.  We also place a conservative upper limit on the
density spectrum inner scale of $l_1 \ll 2000$~km.

We also compared our 1.67~GHz observations to those of \cite{pm81},
acquired approximately 21~years previously.  We find no evidence for
any change in the separation at this frequency.  This places a limit
on the projected linear separation velocity of the two components
of~$c$.

\acknowledgements

The National Radio Astronomy Observatory is a facility of the National Science Foundation operated
under cooperative agreement by Associated Universities, Inc.  Basic
research in radio astronomy at the NRL is supported by the Office of
Naval Research.

\end{document}